\documentclass{desyproc}

\usepackage{amsmath}
\usepackage{amssymb}
\usepackage{mathrsfs}
\usepackage{color}
\usepackage{dcolumn}
\usepackage{graphicx}
\usepackage{multirow}
\usepackage{rotating}
\usepackage{afterpage}
\usepackage{enumerate}
\usepackage{xspace}
\usepackage{url}
\usepackage{booktabs}
\usepackage{wrapfig,rotating}
\usepackage{sidecap}

\usepackage{fancyhdr}
\usepackage{float}
\usepackage{psfrag}
\newcommand{\mm}{\mathrm}

\newcommand{\dzero}     {D\O\xspace}
\newcommand{\wplus}     {$W+$jets\xspace}
\newcommand{\zplus}     {$Z+$jets\xspace}
\newcommand{\muplus}    {$\mu +$jets\xspace}
\newcommand{\eplus}     {$e +$jets\xspace}
\newcommand{\ljets}     {$\ell +$jets\xspace}
\newcommand{\lplustw}     {$\ell +2$\,jets\xspace}
\newcommand{\lplusth}     {$\ell +3$\,jets\xspace}
\newcommand{\lplusfo}     {$\ell +4$\,jets\xspace}
\newcommand{\ttbar}     {$t\bar{t}$\xspace}
\newcommand{\ppbar}     {$p\bar{p}$\xspace}
\newcommand{\met}       {$\not\!\!E_T$\xspace}
\newcommand{\metMath}       {\not\!\!E_T}

\newcommand\T{\rule{0pt}{2.6ex}}       

\begin{document}
\title{Inclusive top pair production at Tevatron and LHC in electron/muon final states}


%

%

%
\author{{\slshape Andreas Jung}  for the ATLAS, CDF, CMS and \dzero Collaboration\\[1ex]
Fermilab, DAB5 - MS 357, P.O. Box 500, Batavia, IL, 60510, USA\\[1ex]
FERMILAB-CONF-13-575-PPD}

\contribID{xy}  
\confID{7095}
\desyproc{DESY-PROC-2013-XY}
\acronym{TOP2013}
\doi            

\maketitle


\begin{abstract}
Recent measurements of the inclusive top pair production at the Tevatron and LHC collider in the electron/muon final states are discussed. Measurements at the Tevatron use up to $9.7$ fb$^{-1}$ of data, and at the LHC up to $4.9$ fb$^{-1}$ of data at $\sqrt{s}=7$ TeV and up to $20.3$ fb$^{-1}$ of data at $\sqrt{s}=8$ TeV. For the experiments at both colliders these corresponds to the full data sets at the given center-of-mass energies. Overall results are in agreement between the experiments at the Tevatron and between the experiments at the LHC. All measurements are in agreement with recent theory calculations at NNLO QCD. Individual LHC measurements are challenging the precision of the theory calculations.
\end{abstract}

\section{Introduction}
The top quark is the heaviest known elementary particle and was discovered at the Tevatron $p\bar{p}$ collider in 1995 by the CDF and \dzero collaboration \cite{top_disc1,top_disc2} with a mass around $173~\mathrm{GeV}$. At the Tevatron the production is dominated by the $q\bar{q}$ annihilation process with 85\% as opposed to gluon-gluon fusion which contributes only 15\%, whereas at the LHC fractions are approximately opposite. The top quark has a very short lifetime, which prevents any hadronization process of the top quark. Instead bare quark properties can be observed by measuring top quark properties.\\
Theoretical predictions of the \ttbar production process exist at various orders of perturbative QCD (pQCD). The most recent prediction is a fully re-summed next-to-next-to-leading log (NNLL) at next-to-next-to-leading order (NNLO) pQCD calculation \cite{nnloTheory}.
\begin {table}[tp]%
\centering %
\begin {tabular}{lccc} 
\toprule %
\multicolumn{1}{l}{Collider} & \multicolumn{3}{c}{} \\
\multicolumn{1}{l}{} & \multicolumn{1}{c}{$\sigma_{\mm{tot}}$ [pb]}
                     & \multicolumn{1}{c}{$\delta_{\mm{scales}}$ [pb]}
                     & \multicolumn{1}{l}{$\delta_{\mm{pdf}}$ [pb]} \\ \midrule
Tevatron ($\sqrt{s}=1.96$~TeV) & $7.164$ & $^{+0.110~(1.5\%)}_{-0.200~(2.8\%)}$ &  $^{+0.169~(2.4\%)}_{-0.122~(1.7\%)}$  \T \\
LHC ($\sqrt{s}=7$~TeV) & 	 $172.0$ & $^{+4.4~~~\,(2.6\%)}_{-5.8~~~\,(3.4\%)}$ &  $^{+4.7~~~\,(2.7\%)}_{-4.8~~~\,(2.8\%)}$  \\
LHC ($\sqrt{s}=8$~TeV) & 	 $245.8$ & $^{+6.2~~~\,(2.5\%)}_{-8.4~~~\,(3.4\%)}$ &  $^{+6.2~~~\,(2.5\%)}_{-6.4~~~\,(2.6\%)}$ \\ \bottomrule
\end {tabular}
\caption {\label{tab:xsecTheory} Total \ttbar production cross sections and their uncertainties \cite{nnloTheory} at the Tevatron and the LHC.}
\end {table}
Table \ref{tab:xsecTheory} summarizes these predictions for the Tevatron and the LHC center-of-mass energies (using $m_t = 173$ GeV and the MSTW2008NNLO PDF). The total uncertainty from factorization and renormalization scale variation and PDF uncertainties is approximately 3.5\% for the Tevatron and approximately 4.3\% at the LHC.

\subsection{Measurement of cross sections}
The measurements presented here are performed using either the dilepton ($\ell \ell$) final state or the lepton+jets (\ljets) final state, where $\ell$ can be an electron or a muon (details on the lepton reconstruction can be found here \cite{bquarkID_top2013}). The branching fraction for top quarks decaying into $Wb$ is almost 100\%. Within the \ljets~final state one of the $W$ bosons (stemming from the decay of the top quarks) decays leptonically, the other $W$ boson decays hadronically. For the dilepton final state both $W$ bosons decay leptonically. The main background contribution in the \ljets~decay channel originates from \wplus production, whereas the dilepton decay channel suffers most from contributions from \zplus production. At the LHC also single top quark production is one of the dominant background contributions for the $\ell \ell$ and \ljets channel. Jets originating from a beauty quark ($b$-jets) are usually identified by means of multivariate discriminants built by the combination of variables describing the properties of secondary vertices and of tracks with large impact parameters relative to the primary vertex.\\
The measured cross section can be calculated by using
\begin{equation}
 \label{eqn:xsecDef}
\sigma_{\mm{tot}}(t\bar{t}) = \dfrac{N^{\mm{obs}} - N^{\mm{bg}}}{\epsilon \cdot \mathcal{A} \cdot {\mathscr{L}} \cdot {\mathscr{B}}}~.
\end{equation}
The number of observed data events $N^{\mm{obs}}$ is subtracted by the number of expected background events $N^{\mm{bg}}$ and then corrected for the detector efficiency $\epsilon$ and acceptance $\mathcal{A}$, the total integrated luminosity $\mathscr{L}$ that corresponds to the selection requirements, and for the branching fraction $\mathscr{B}$ into the decay channel under consideration. Thus any measurement of a cross section relies on Monte-Carlo (MC) samples to correct the data for the detector efficiency and also in order to extrapolate from the fiducial cross section to the total cross section. For this purpose all cross section measurements (also differential) use currently theory predictions at leading-order or next-to-leading order pQCD.

%

\section{Dilepton channel (CDF)}
CDF uses all available data corresponding to $8.8~\mathrm{fb^{-1}}$ in the dilepton decay channel to measure the \ttbar production cross section \cite{cdf_dilepton}. The data is selected by requiring exactly two leptons and the accompanying missing transverse energy \met originating from the non-reconstructed neutrinos from the leptonic decays of the two $W$ bosons. 
\begin{SCfigure}
  \centering
\includegraphics[width=0.60\textwidth]{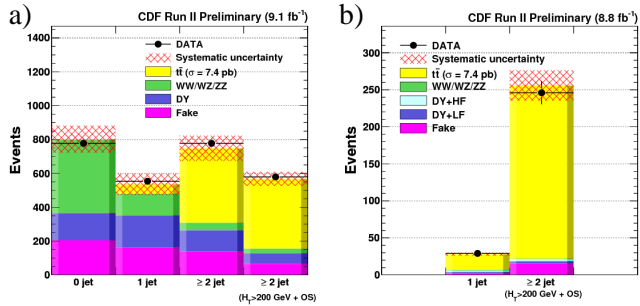}
\caption{\label{Fig:cdf_dilepton} Selected data in the dilepton channel for the a) pre-tag or b) $b$-tagged case compared to the background contributions.}
\end{SCfigure}
Leptonic decays of $\tau$ are included, whereas hadronic ones are not considered here. At least one isolated electron with $E_T > 20$ GeV is required, whereas the second electron does not need to be isolated. Muons are required to have at least $p_T > 20$ GeV, again at least one muon needs to be isolated. Furthermore at least two jets with $p_T > 15$ GeV and pseudorapidity $|\eta| < 2.5$ are required, with at least one jet identified to originate from a $b$-quark ($b$-tagged). Figure \ref{Fig:cdf_dilepton} shows the selected data in the a) pre-tag or b) $b$-tagged case. The total cross section (see Equation \ref{eqn:xsecDef}) assuming $m_t = 172.5$ GeV is measured from this $b$-tagged event selection to $\sigma_{\mm{tot}}(t\bar{t}) = 7.09 \pm 0.49 (\mm{stat}) \pm 0.52 (\mm{sys}) \pm 0.43 (\mm{lumi})$ pb. The systematic uncertainty is dominated by the modeling of the $b$-tagging and the total uncertainty for this measurement is 12\%. If no requirement on $b$-tagging is applied a cross section of $\sigma_{\mm{tot}}(t\bar{t}) = 7.66 \pm 0.44 (\mm{stat}) \pm 0.52 (\mm{sys}) \pm 0.47 (\mm{lumi})$ pb is measured. Table \ref{tab:xsecTevatronSummary} shows a comparison to other CDF measurements in the \ljets channel not presented at TOP2013 \cite{cdf_ljetsann,cdf_ljetssvx} and to \dzero measurements (see Sec.\ \ref{tab:d0results}). The measurements are in good agreement with the most recent pQCD prediction at NNLO, which yields a cross section of $\sigma_{\mm{tot}}(t\bar{t}) = 7.24 ^{+0.23}_{-0.27} (\mathrm{scales} \oplus \mathrm{pdf})$ pb.

\section{Dilepton and \ljets channel (\dzero)}
\label{tab:d0results}
In case of \dzero three recent measurements of the \ttbar cross section have been presented. The measurement in the dilepton channel corresponds to 5.4 fb$^{-1}$ of integrated luminosity \cite{d0_dilepton}. Events are required to have two isolated leptons with $p_T > 15$ GeV, \met and at least one (two) jet with $p_T > 20$ GeV in the $e\mu$ ($ee$, $\mu \mu$) channel. Further cuts are applied to improve signal purity and reject background contributions in four different categories: $ee$ and $\mu\mu$ with each at least two jets, $e\mu + $ 1 jet and $e\mu + $ 2 jets. The discriminant distribution for identifying jets stemming from $b$ quarks in the four event categories is used to maximize a likelihood function. 
\begin{figure}[h]
\centering
\includegraphics[width=0.25\textwidth]{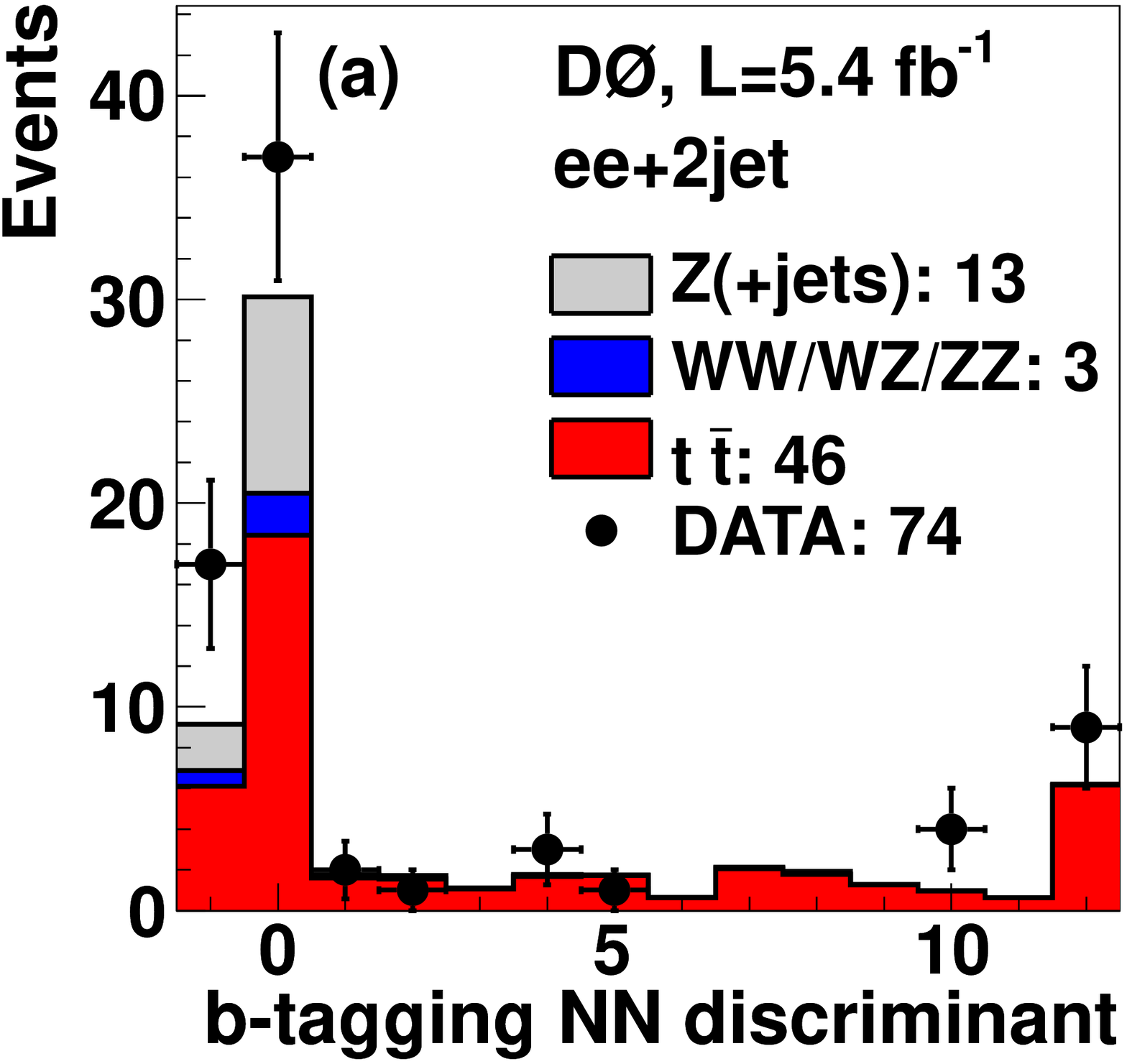}
\includegraphics[width=0.3675\textwidth]{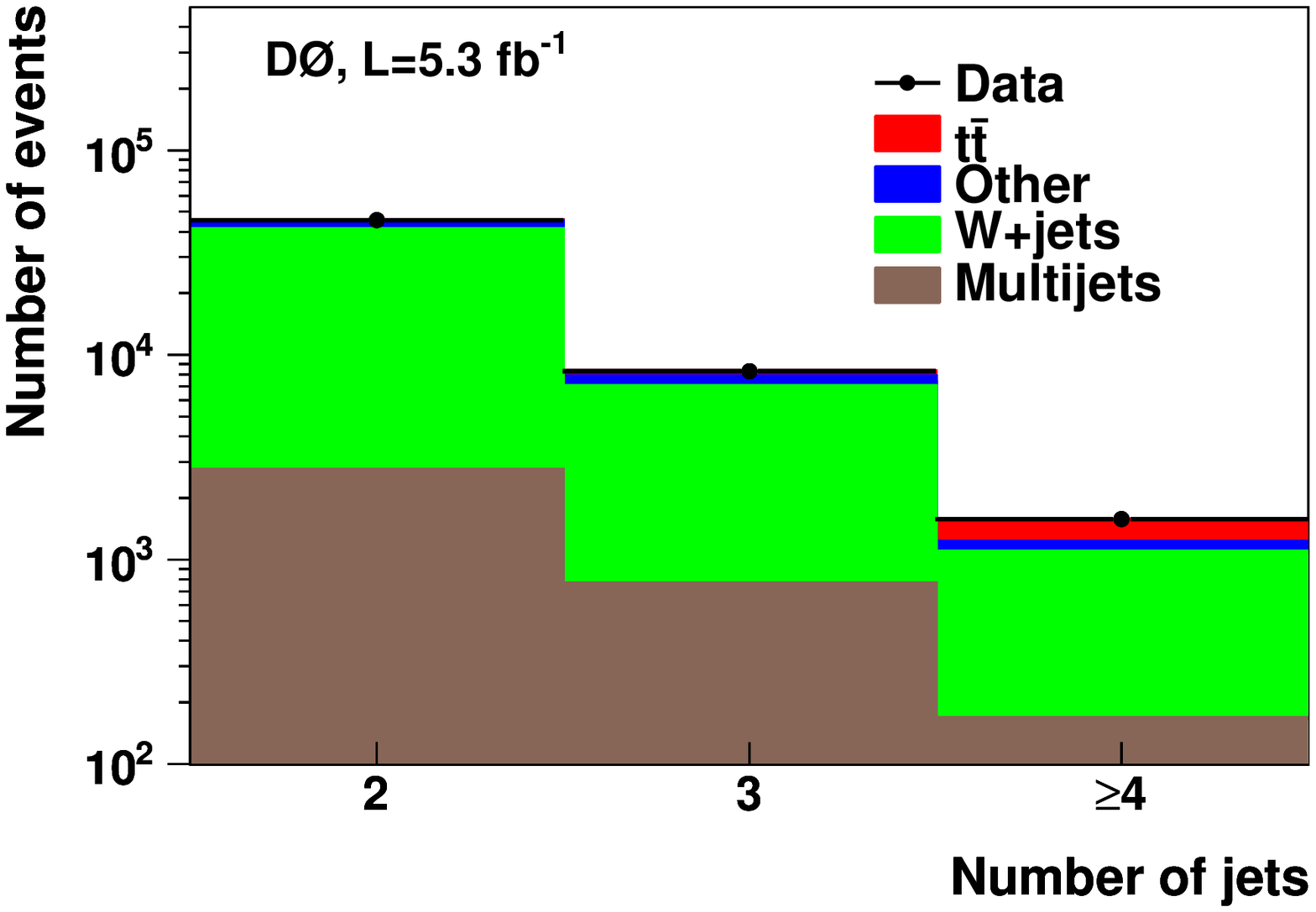}
\includegraphics[width=0.3675\textwidth]{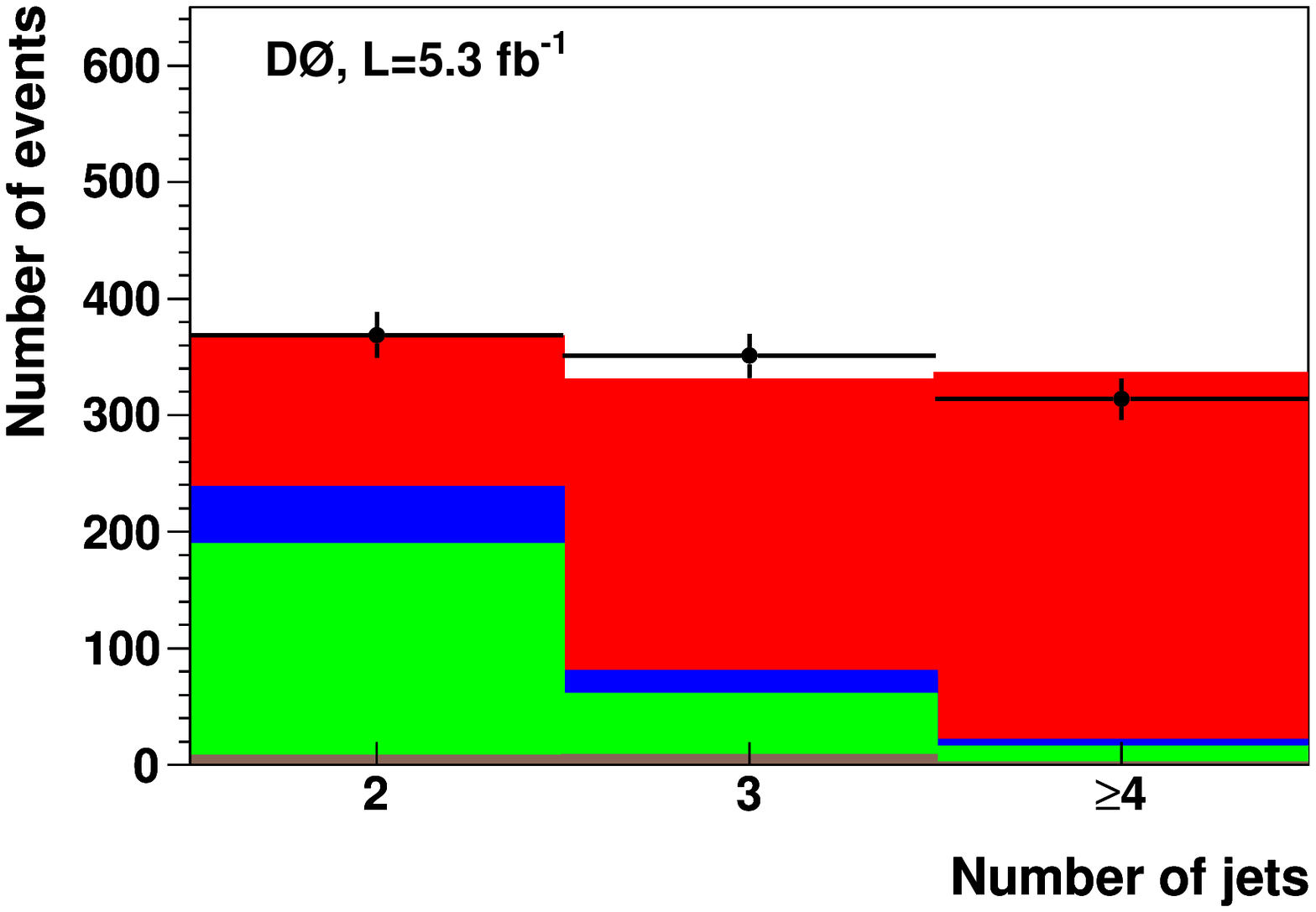}
\caption{\label{fig:d0_dilepton} The $b$-tagging discriminant output is shown in a) for the $ee + 2$ jets sample, where the expected \ttbar cross section is normalized to $7.45$ pb. The sample composition of the selected \ljets data as a function of the jet multiplicity for b) zero $b$-tagged jets and c) more than $1$ $b$-tagged jet.}
\end{figure}%
An example $b$-tagging discriminant distribution is shown in Figure \ref{fig:d0_dilepton} a), where the expected \ttbar cross section is normalized to $7.45$ pb. The cross section is measured to be $\sigma_{\mm{tot}}(t\bar{t}) = 7.36 ^{+0.90}_{-0.76} (\mm{stat} + \mathrm{sys})$ pb.\\
The measurement in the \ljets channel uses $5.4$ fb$^{-1}$ of data \cite{d0_ljets}. Events compatible with the \ljets signature are selected by requiring at least 2 jets with $p_T > 20$ GeV and within $|\eta|<2.5$, in addition an isolated lepton with $p_T > 20$ GeV is required. Electrons are selected within $|\eta|<1.1$, whereas for muons $|\eta|<2.0$ is required. Additionally \met$ > 20\,(25)$ GeV is required in the \eplus (\muplus) channel. The selected data are used for a combined measurement using $b$-tagging and kinematic information splitted in different channels by number of $b$-tags and jets. In addition to the \ttbar cross section the heavy flavor contribution is fitted as well. 
Figure \ref{fig:d0_dilepton} b) and c) shows the sample composition of the selected data as a function of the jet multiplicity for $0$ and more than $1$ $b$-tagged jet. As one expects the background contributions rise towards lower jet multiplicity and the \ttbar contribution rises strongly with number of $b$-tags (and also with number of jets). The cross section is measured to be $\sigma_{\mm{tot}}(t\bar{t}) = 7.78 \pm 0.25(\mm{stat}) \pm ^{0.65}_{0.58}(\mm{sys + lumi})$ pb, which is in good agreement to the theory prediction and other measurements by \dzero or CDF.\\
The most recent measurement using \ljets events selected in the full data set is derived from differential top quark cross section distributions \cite{d0note_diff}. The measurement is not optimized to measure the total cross section, hence it suffers from a larger total uncertainty. A dedicated inclusive \ttbar cross section measurement in \ljets channel is in progress. For the presented measurement \ljets events are selected by requiring an isolated lepton $(e/\mu)$ with $p_T > 20$ GeV, \met $> 20$ GeV and at least four jets with $p_T > 20$ GeV and $|\eta| < 2.5$. Further cuts are applied to improve data quality and reject background \cite{d0note_diff}. To increase the signal purity at least one $b$-tagged jet is required. The sample composition is determined using data in the \lplustw, \lplusth and inclusive \lplusfo bin and a fit to the discriminant distribution of the $b$-tagging. 
\begin{figure}[hb]
\centerline{
\includegraphics[width=0.475\textwidth]{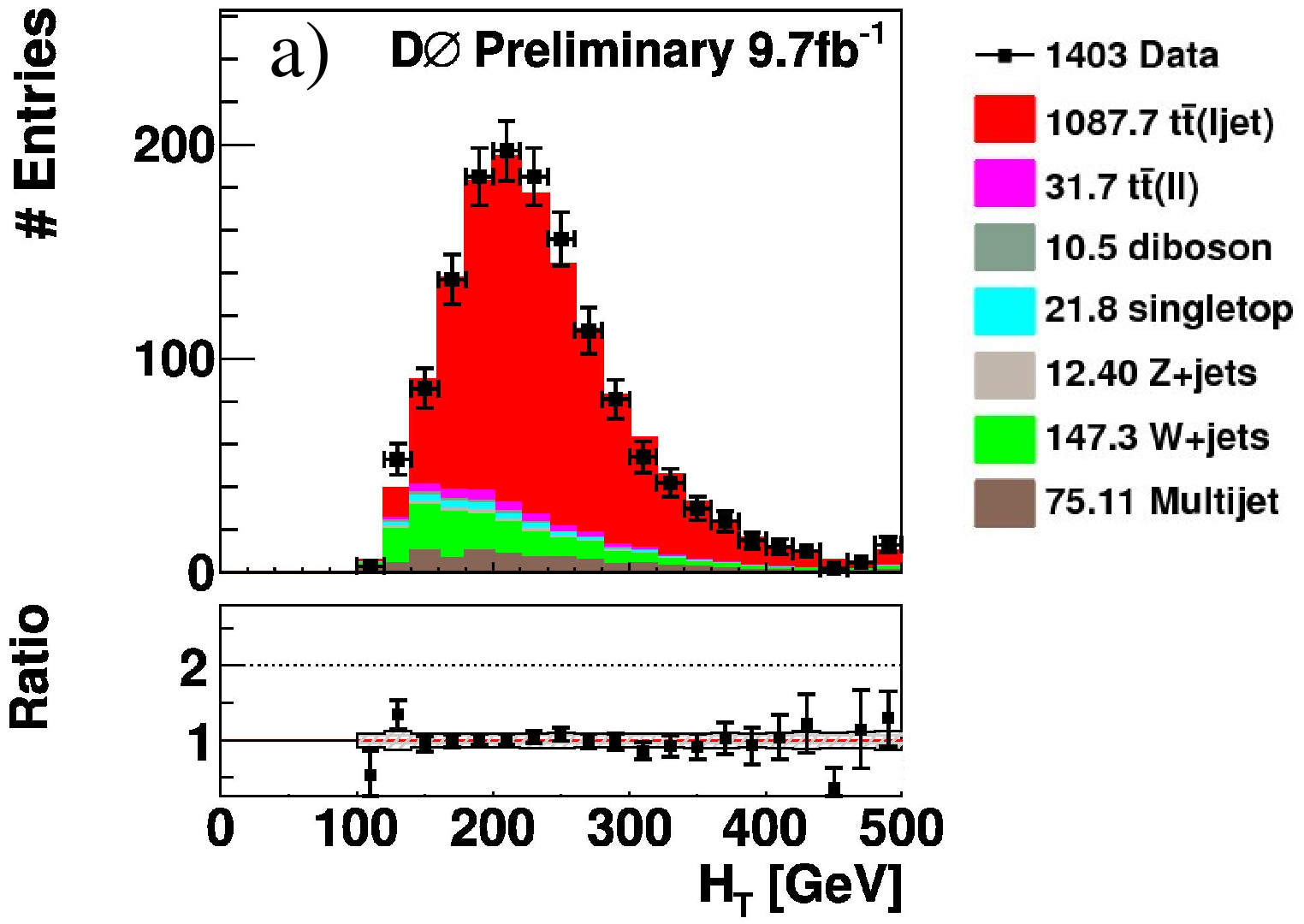}
\includegraphics[width=0.475\textwidth]{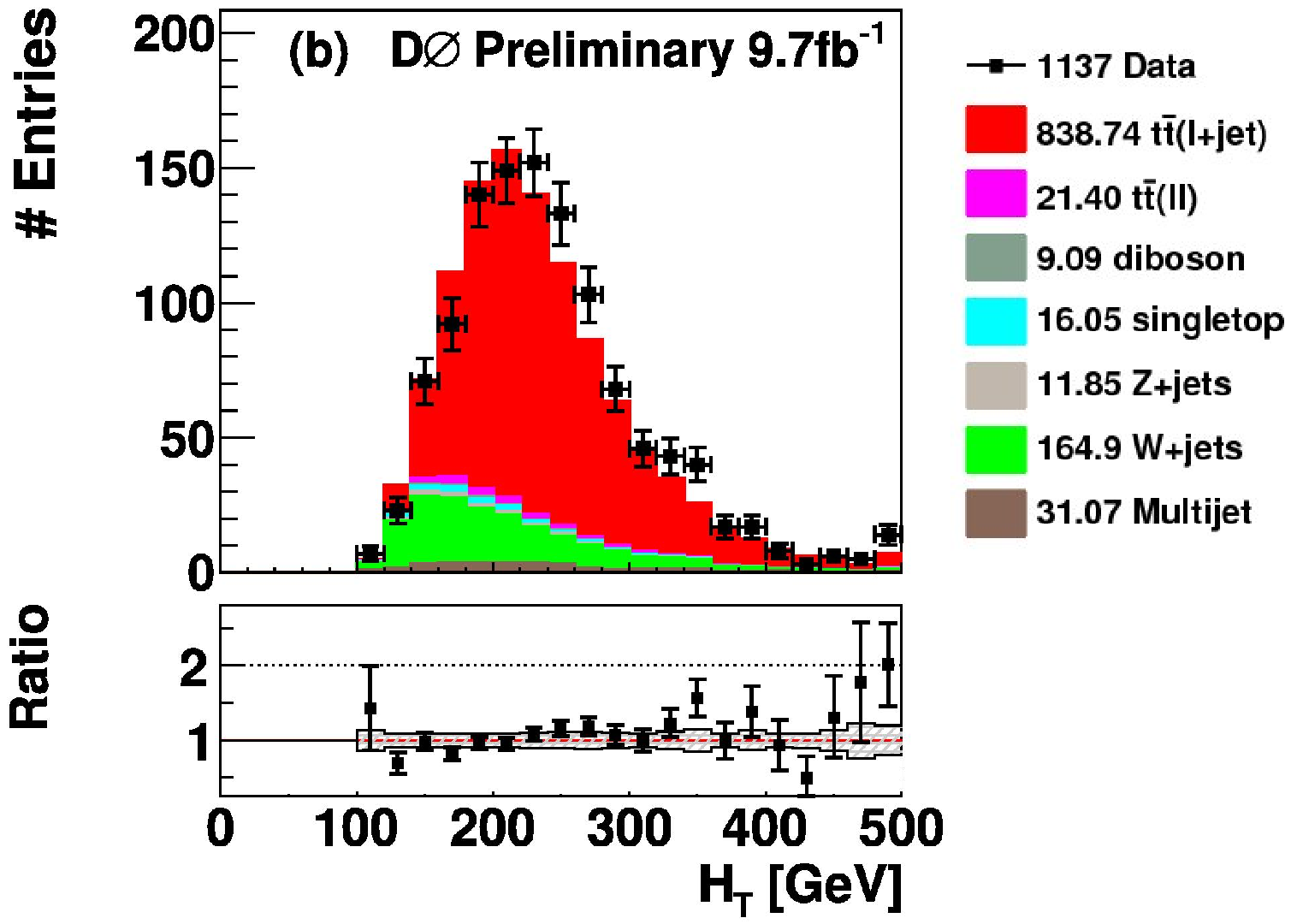}
}
\caption{\label{fig:ljets97_control} The scalar sum $H_T$ of the $p_T$ of the jets for the a) \eplus and b) \muplus channel, the expected \ttbar cross section is normalized to the measured cross section of 8.27 pb.}
\end{figure}
The level of agreement between data and MC is good and can be seen in Figure \ref{fig:ljets97_control} a) and b) for the scalar sum $H_T$ of the $p_T$ of the jets for the \eplus and \muplus channel, respectively. Thus the established sample composition in the \lplusfo bin is used to measure differential \ttbar cross section by first performing a kinematic fit to identify the top quarks. The output is subtracted by the background and the cross section is determined according to equation \ref{eqn:xsecDef}, where data is corrected by means of a regularized matrix unfolding. The total \ttbar cross section is then measured to be $\sigma_{\mm{tot}}(t\bar{t}) = 8.27 \pm 0.68(\mm{stat}) \pm ^{0.61}_{0.58}(\mm{sys}) \pm 0.50 (\mm{lumi})$ pb, which is somewhat higher than the SM prediction but given the uncertainties still in agreement.\\
A comparison of all the discussed measurements at the Tevatron is presented in Table \ref{tab:xsecTevatronSummary}. More information about other Tevatron top quark measurements can be found here \cite{Tevwebpages}.
\begin {table}[tp]%
\centering %
\begin {tabular}{lclc}
\toprule %
Measurement & ${\mathscr{L}}$ [fb$^{-1}$] & $\sigma_{\mm{tot}}(t\bar{t})$ [pb] & total rel.\,unc. \\ \midrule
CDF ($\ell \ell$, $b$-tag) & 8.8 & $7.09 \pm 0.49 (\mm{stat}) \pm 0.52 (\mm{sys}) \pm 0.43 (\mm{lumi})$ & 12\% \T \\
CDF (\ljets \cite{cdf_ljetsann}) & 4.6 & $7.82 \pm 0.38(\mm{stat}) \pm 0.37(\mm{sys}) \pm 0.15 (\sigma_Z)$ & 7.0\%\\
CDF (\ljets \cite{cdf_ljetssvx}) & 4.3 & $7.32 \pm 0.35(\mm{stat}) \pm 0.59(\mm{sys}) \pm 0.15 (\sigma_Z)$ & 9.6\%\\
\dzero ($\ell \ell$) & 5.4 & $7.36 ^{+0.90}_{-0.76} (\mm{stat} \oplus \mathrm{sys})$ & 11\% \\
\dzero (\ljets, $b$-tag) & 9.7 & $8.27 \pm 0.68(\mm{stat}) \pm ^{0.61}_{0.58}(\mm{sys}) \pm 0.50 (\mm{lumi})$ & 12.5\%\\
\dzero (\ljets, $b$-tag) & 5.3 & $7.78 \pm 0.25(\mm{stat}) \pm ^{0.65}_{0.58}(\mm{sys \oplus lumi})$ & 9.1\%\\ \\
Theory: & & & \\
NNLO pQCD \cite{nnloTheory} & NA & $7.24 ^{+0.23}_{-0.27} (\mathrm{scales} \oplus \mathrm{pdf})$ & 3.5\% \\ \bottomrule
\end {tabular}
\caption {\label{tab:xsecTevatronSummary} Summary of presented and discussed measurements of the total \ttbar production cross sections and their uncertainties at the Tevatron. The \dzero measurement in \ljets using 9.7 fb$^{-1}$ is not optimized to measure the total cross section, hence the larger total uncertainty.}
\end {table}
The uncertainties of a single measurement at the Tevatron are significantly larger than the uncertainties of the most current pQCD calculation ($\approx 3.5$\%), and only the combination of all available Tevatron cross section measurements \cite{TevatronCombis} yields an uncertainty closer to the theoretical one.

\section{Dilepton and \ljets channel (ATLAS)}
The production of \ttbar pairs at the LHC is strongly enhanced by the higher energy, resulting in $20-30$ times higher cross sections if compared to the Tevatron. Thus measurements in the dilepton and \ljets channel are not statistically limited and the uncertainties are dominated by systematic uncertainties. Both channels can be selected with high purity, with the dilepton $e\mu$ channel being almost background free. \\
ATLAS performed a measurement of the \ttbar cross section in the \ljets channel using $5.8$ fb$^{-1}$ of data at $\sqrt{s} = 8$ TeV \cite{atlas_ljets8tev}. Events are triggered by the single electron or muon trigger and verified off-line by requiring a reconstructed isolated $e$ or $\mu$ with $p_T >40$ GeV and $|\eta|< 2.5$, electrons in the region $1.37 < |\eta|< 1.52$ are excluded. A second lepton is vetoed. For electrons (muons) a cut of \met$> 30\,(20)$ GeV, as well as a cut on the transverse mass of the $W$ of $M_T(W) > 30$ GeV is required. For muons the sum of $M_T(W) + \metMath$ is required to be larger than 60 GeV. Furthermore at least three jets with $p_T > 25$ GeV and $|\eta|<2.5$ are required, and at least one of the jets needs to be $b$-tagged. Additional quality cuts are applied \cite{atlas_ljets8tev}. 
\begin{SCfigure}[][h]
  \centering
  \includegraphics[width=0.285\textwidth]{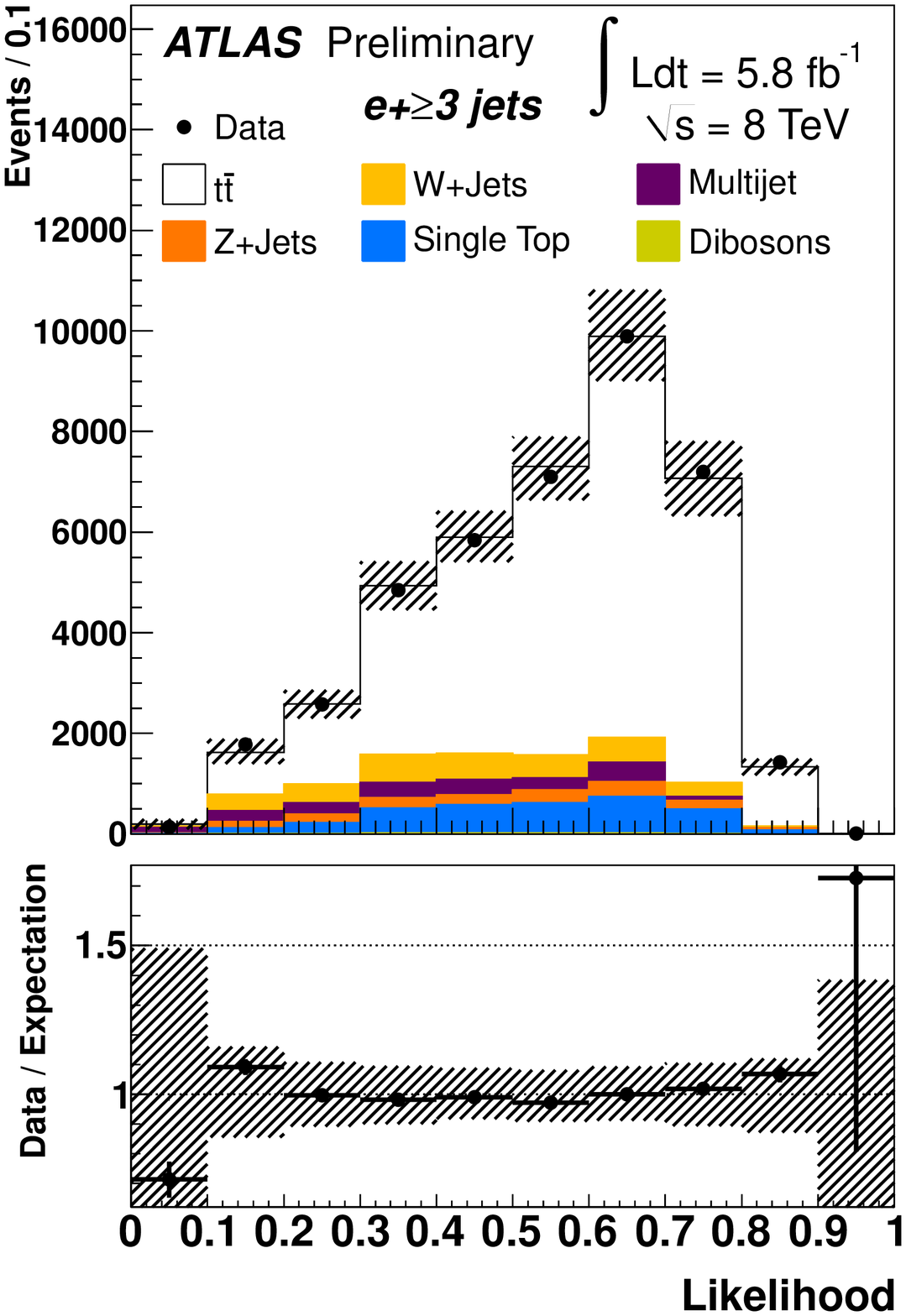}
  \hspace{15pt}
  \includegraphics[width=0.285\textwidth]{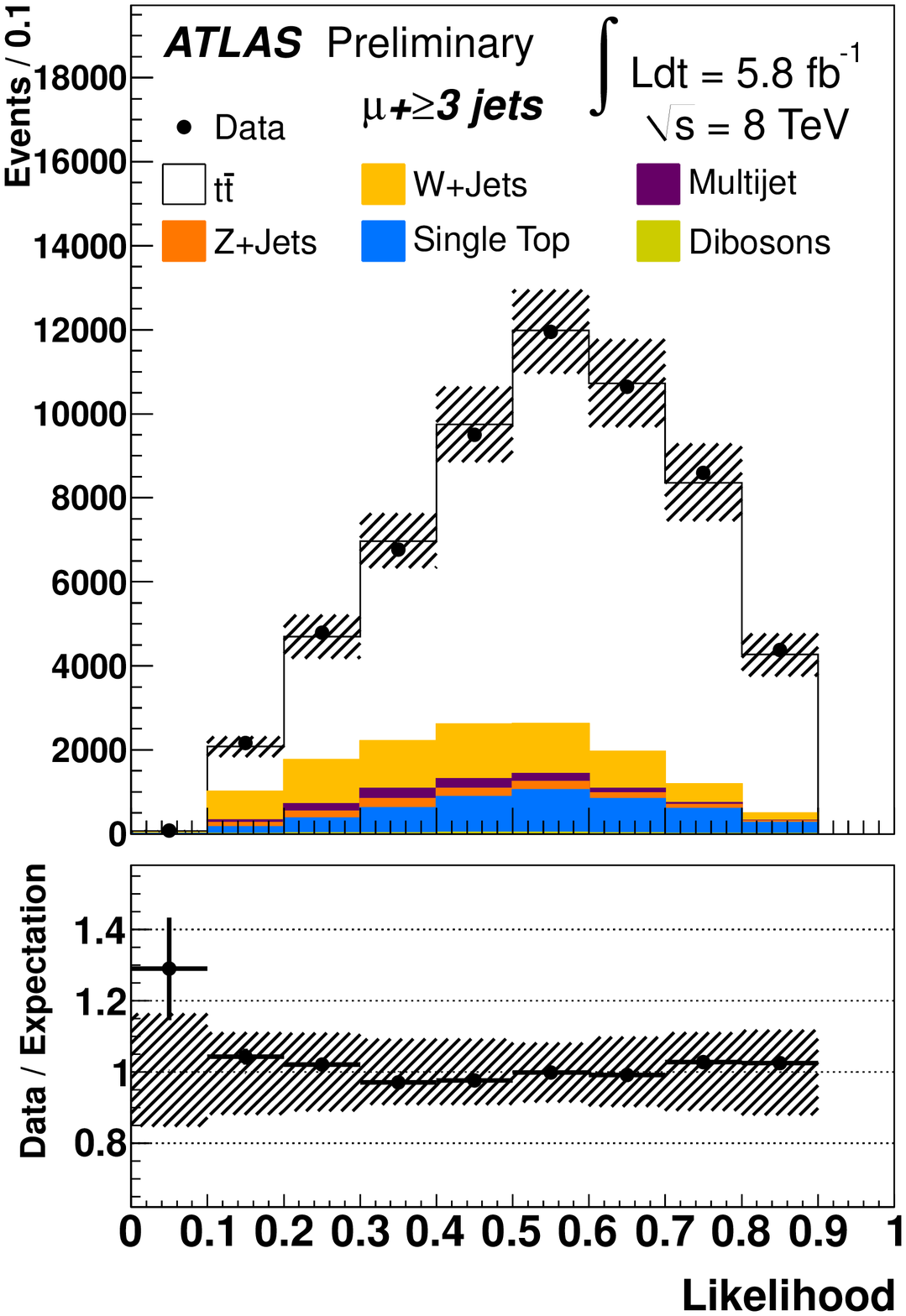}
  \caption{\label{fig:atlas_ljetsLikeli} The likelihood discriminant based on lepton $\eta$ and the aplanarity $A'$ is shown in a) for the \eplus and in b) for the \muplus channel. The expected \ttbar cross section is normalized to the measured cross section.}
\end{SCfigure}
The \ttbar cross section is measured from a fit to a likelihood discriminant $D_i = L_i^s / (L_i^s + L_i^b)$ based on the lepton $\eta$ and the transformed event aplanarity $A'$, which is given by $A' = \exp(-8\cdot A)$ (the aplanarity $A$ is based on the momenta of all jets and the lepton). Figure \ref{fig:atlas_ljetsLikeli} shows the likelihood discriminant for the a) \eplus channel and b) for the \muplus channel. Both channels are well described by the background contributions and in contrast to \ttbar production in \ppbar collisions, single top quark production is amongst the dominant background contributions. The total \ttbar cross section assuming $m_t = 172.5$ GeV is then measured to be $\sigma_{\mm{tot}}(t\bar{t}) = 241 \pm 2(\mm{stat}) \pm 31 (\mm{sys}) \pm 9 (\mm{lumi})$ pb. The dominant systematic uncertainties arise from the modeling of the signal implemented in MC, the jet and \met reconstruction and calibration followed by the lepton trigger, identification and reconstruction. The measurement agrees with CMS measurements (see Table \ref{tab:xsecLHCSummary}) and is in agreement with the latest theory prediction of $\sigma_{\mm{tot}}(t\bar{t}) = 252.9 ^{+13.3}_{-14.5} (\mm{scales \oplus pdf \oplus \alpha_s})$ pb \cite{nnloTheory}.

The ATLAS measurement in the dilepton channel ($e\mu$) uses the full data set available at $\sqrt{s} = 8$ TeV \cite{atlas_dilepton}. It is a sample with very high purity comprising $\mathcal{O}(10^4)$ \ttbar pairs. Events are selected by the single electron or muon trigger and verified off-line by requiring a reconstructed isolated lepton with $p_T > 25$ and $|\eta| < 2.5$. At least one jet with $p_T > 25$ and $|\eta| < 2.5$ is required and exactly one or exactly two of the jets needs to be $b$-tagged. \vspace{75pt}
\begin{figure}[hb]
\centerline{
\hspace{-75pt}
\includegraphics[width=0.475\textwidth, angle = 270,scale=0.45]{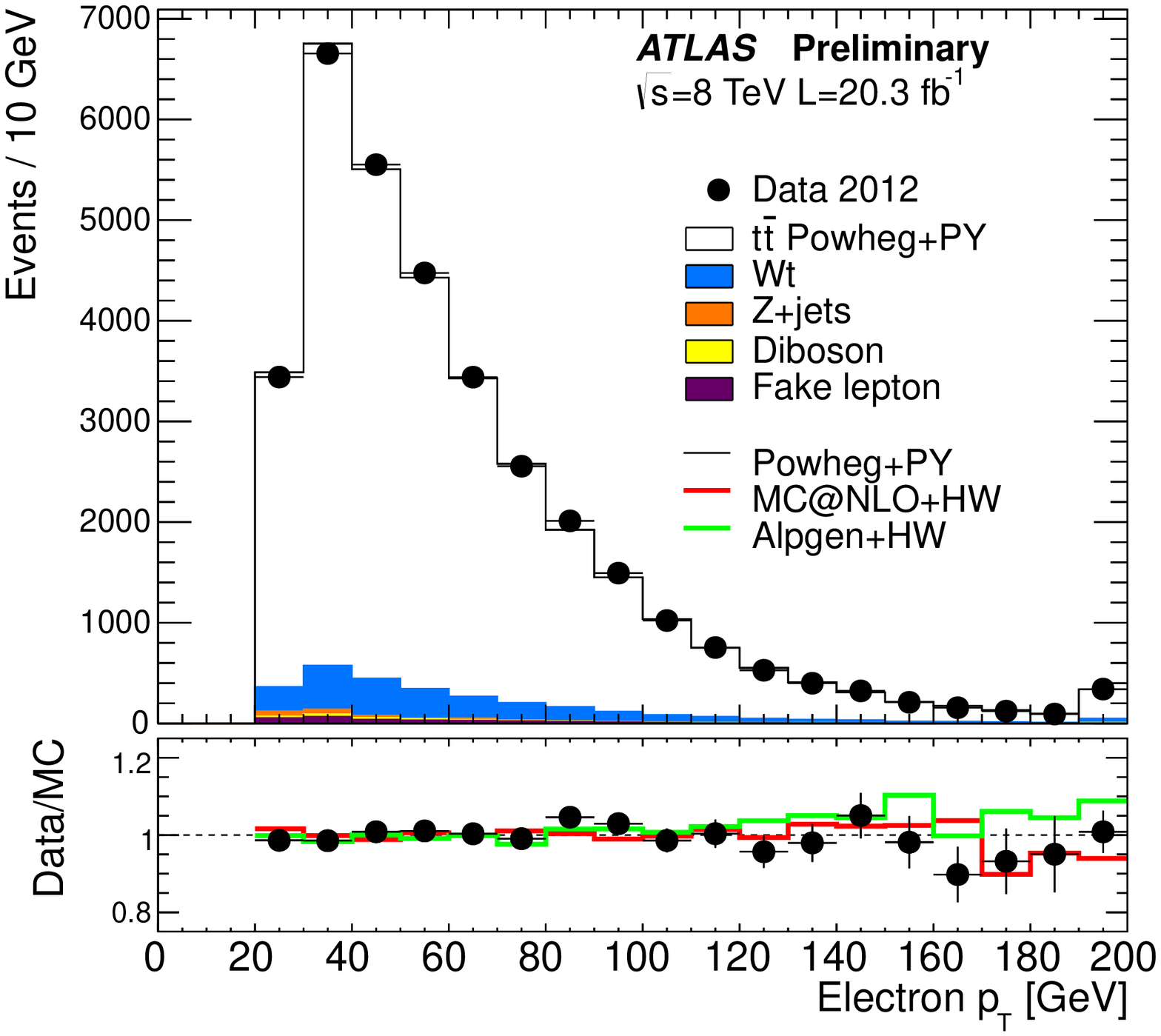}
\hspace{-75pt}
\includegraphics[width=0.475\textwidth, angle = 270,scale=0.45]{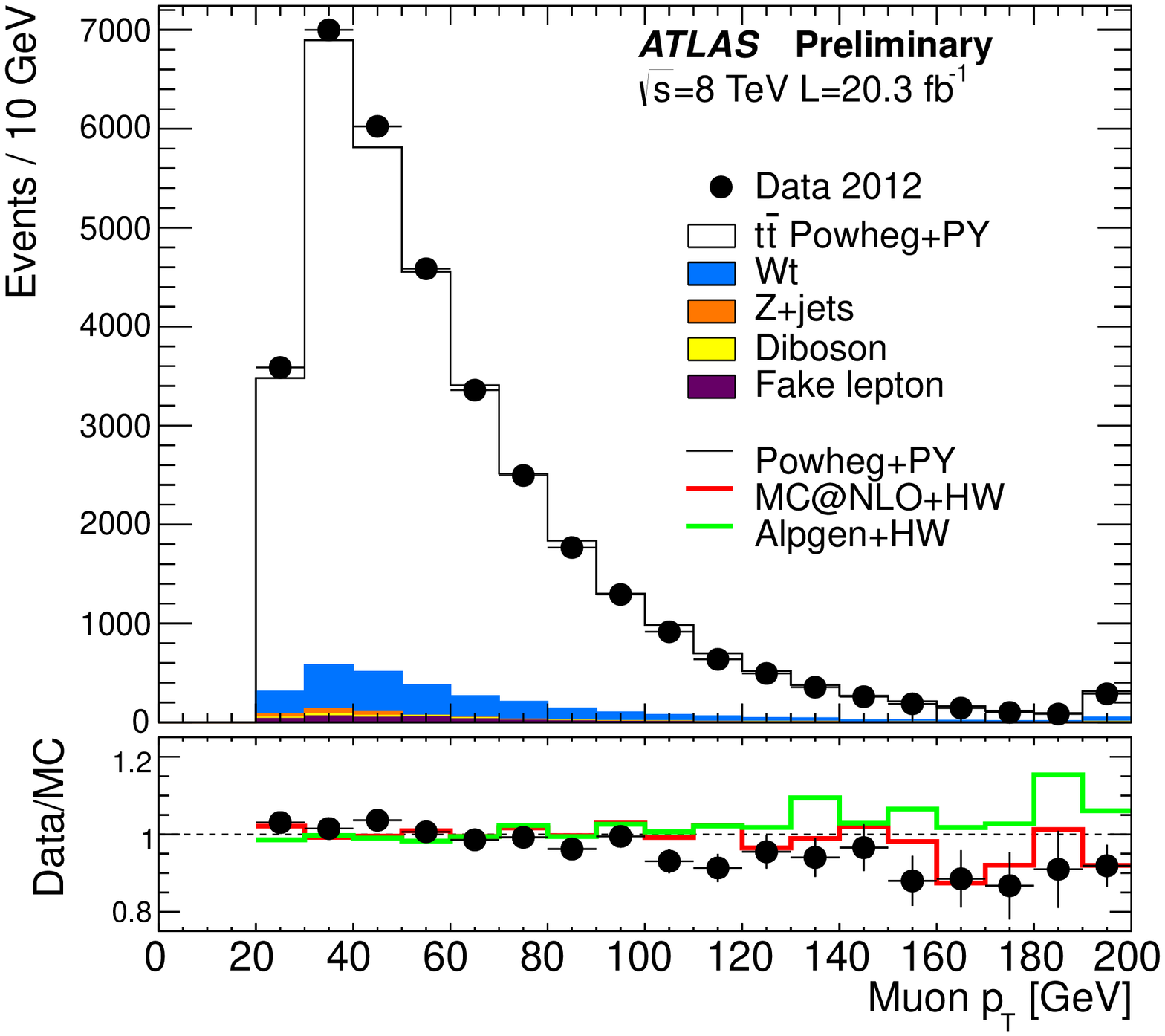}
}
\caption{\label{fig:atlas_dilepton} Comparison of data to signal and background contributions as a function of the lepton $p_T$ for the a) electron and for the b) muon.}
\end{figure}
The level of agreement between data and MC is very good and shown in Figure \ref{fig:atlas_dilepton} a) for the electron $p_T$ and in b) for the muon $p_T$. The \ttbar cross section is measured from a simultaneous determination of $\sigma_{\mm{tot}}(t\bar{t})$ and the efficiency to reconstruct and $b$-tag jets. The measurement strategy is aimed to reduce the related systematic uncertainties of jets and $b$-tagging. The samples with exactly one $b$-jet ($N_1$) and two $b$-jets ($N_2$) are simultaneously fitted with \mbox{$N_1 = \mathscr{L} \cdot \sigma_{t\bar{t}} \cdot \epsilon_{e\mu} \cdot 2\epsilon_b (1-C_b\epsilon_b) + N_1^{\mm{bg}}$} and \mbox{$N_2 = \mathscr{L} \cdot \sigma_{t\bar{t}} \cdot \epsilon_{e\mu} \cdot C_b \epsilon_b^2 + N_1^{\mm{bg}}$}, where $\epsilon_{e\mu}$ is the efficiency to pass the $e\mu$ preselection and $\epsilon_b$ is the combined probability for a jet from the $t \rightarrow Wq$ to be within acceptance, reconstructed as jet and $b$-tagged. The correlation between two $b$-tagged jets is taken into account by $C_b$. Employing this new approach and assuming $m_t = 172.5$ GeV yields a total \ttbar cross section of $\sigma_{\mm{tot}}(t\bar{t}) = 237.7 \pm 1.7(\mm{stat}) \pm 7.4 (\mm{sys}) \pm 7.4 (\mm{lumi}) \pm 4.0 (\mm{beam~energy})$ pb. The latter uncertainty is estimated from a $0.66$\% uncertainty on the beam energy, which translates to a 1.7\% uncertainty on the \ttbar cross section. The combined fit of the \ttbar cross section, the lepton, and the jet efficiencies (including $b$-tagging) significantly reduces the systematic uncertainties from these sources. Thus the measurement is dominated by signal model, the electron identification and the PDF uncertainties. The measurement is in agreement with the latest theory prediction at NNLO pQCD, which yields $\sigma_{\mm{tot}}(t\bar{t}) = 252.9 ^{+13.3}_{-14.5} (\mm{scales \oplus pdf \oplus \alpha_s})$ pb \cite{nnloTheory}.

\section{Dilepton and \ljets channel (CMS)}
CMS performed a combined measurement of the heavy flavor contribution and the
\begin{wrapfigure}{r}{0.50\textwidth}
\centerline{\includegraphics[width=0.47\textwidth]{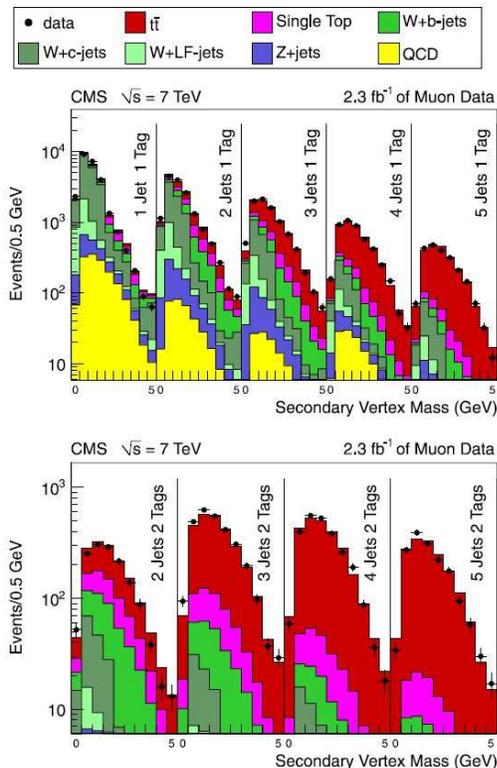}}
\caption{\label{fig:cmsljets7tev} The secondary vertex mass distribution for different bins of the number of jets and $b$-tags. The expected \ttbar cross section is normalized to $7.45$ pb.}
\end{wrapfigure}%
\ttbar cross section in the \ljets decay channel using $2.3$ fb$^{-1}$ of data at $\sqrt{s} = 7$ TeV \cite{cms_ljets7tev}. Events are triggered by the single $e$- or $\mu$-trigger and the off-line selection requires an isolated lepton with $p_T > 35$ GeV. Additionally electrons (muons) are required to have $|\eta| < 2.5\,(2.1)$. The \met originating from the non-reconstructed neutrinos is required to be larger than $20$ GeV. At least one jet with $p_T > 35$ GeV and $|\eta| < 2.4$ is required and at least one of the jets has to be identified as a $b$-jet. As an example Figure \ref{fig:cmsljets7tev} shows the distribution of the secondary vertex mass for the \muplus channel. A maximum likelihood fit using the secondary vertex mass distribution in different number of jet and $b$-tag bins is performed. The total \ttbar cross section assuming $m_t = 172.5$ GeV is then measured to be $\sigma_{\mm{tot}}(t\bar{t}) = 158.1 \pm 2.1(\mm{stat}) \pm 10.2 (\mm{sys}) \pm 3.5 (\mm{lumi})$ pb, which is in agreement with the latest theory prediction of $\sigma_{\mm{tot}}(t\bar{t}) = 172 ^{+4.4}_{-5.8} (\mm{scales})~^{+4.7}_{-4.8} (\mm{pdf})$ pb \cite{nnloTheory}. Table \ref{tab:xsecLHCSummary} gives a comparison to an ATLAS measurement in the \ljets channel \cite{atlas_ljets7} as well as to results in the dilepton channel \cite{atlas_dilepton7, cms_dilepton7} (these result were not presented at TOP2013).\\

\begin{SCfigure}[][ht]
\centering
\includegraphics[angle = 270,width=0.35\textwidth]{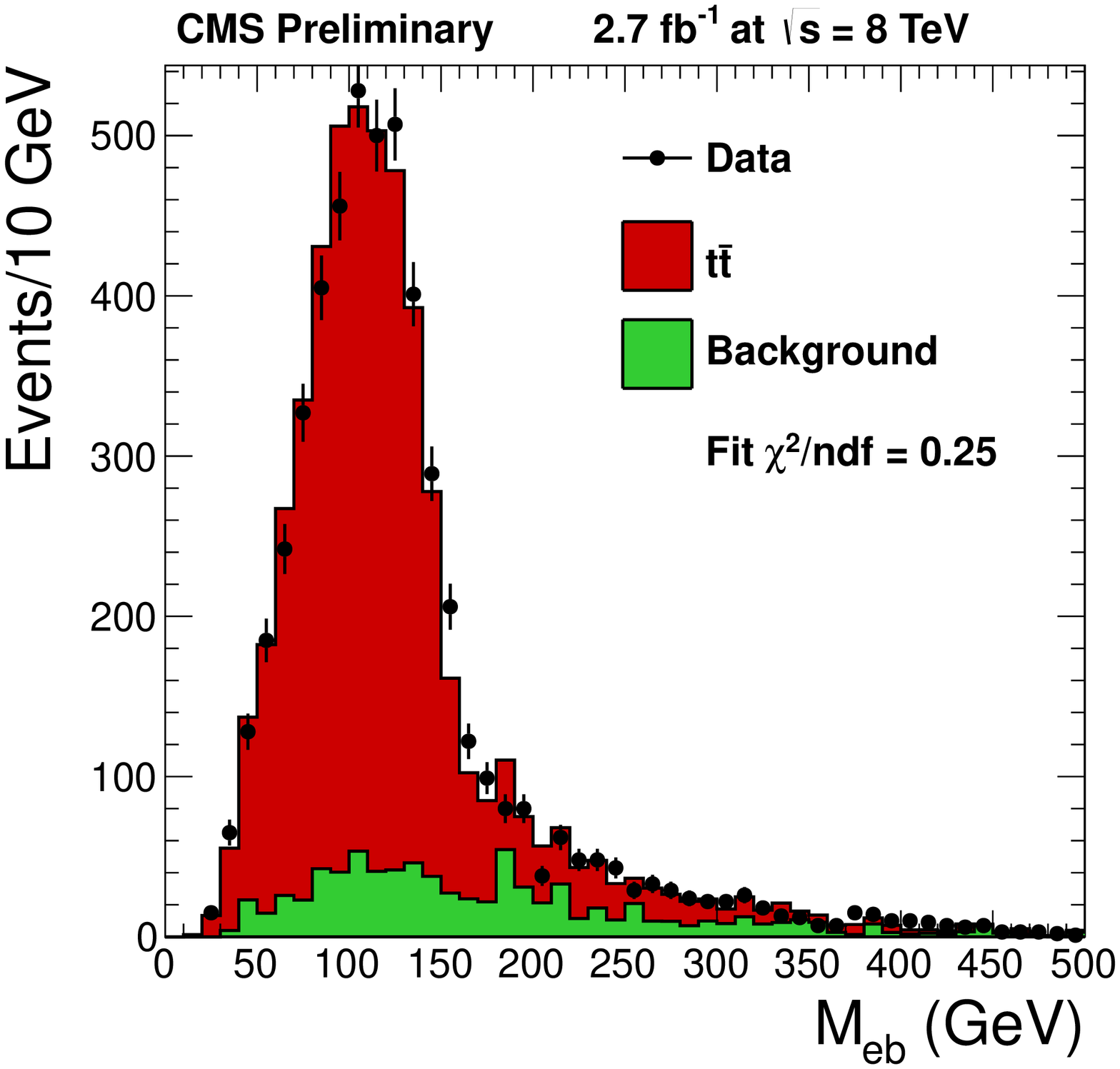}
\includegraphics[angle = 270,width=0.35\textwidth]{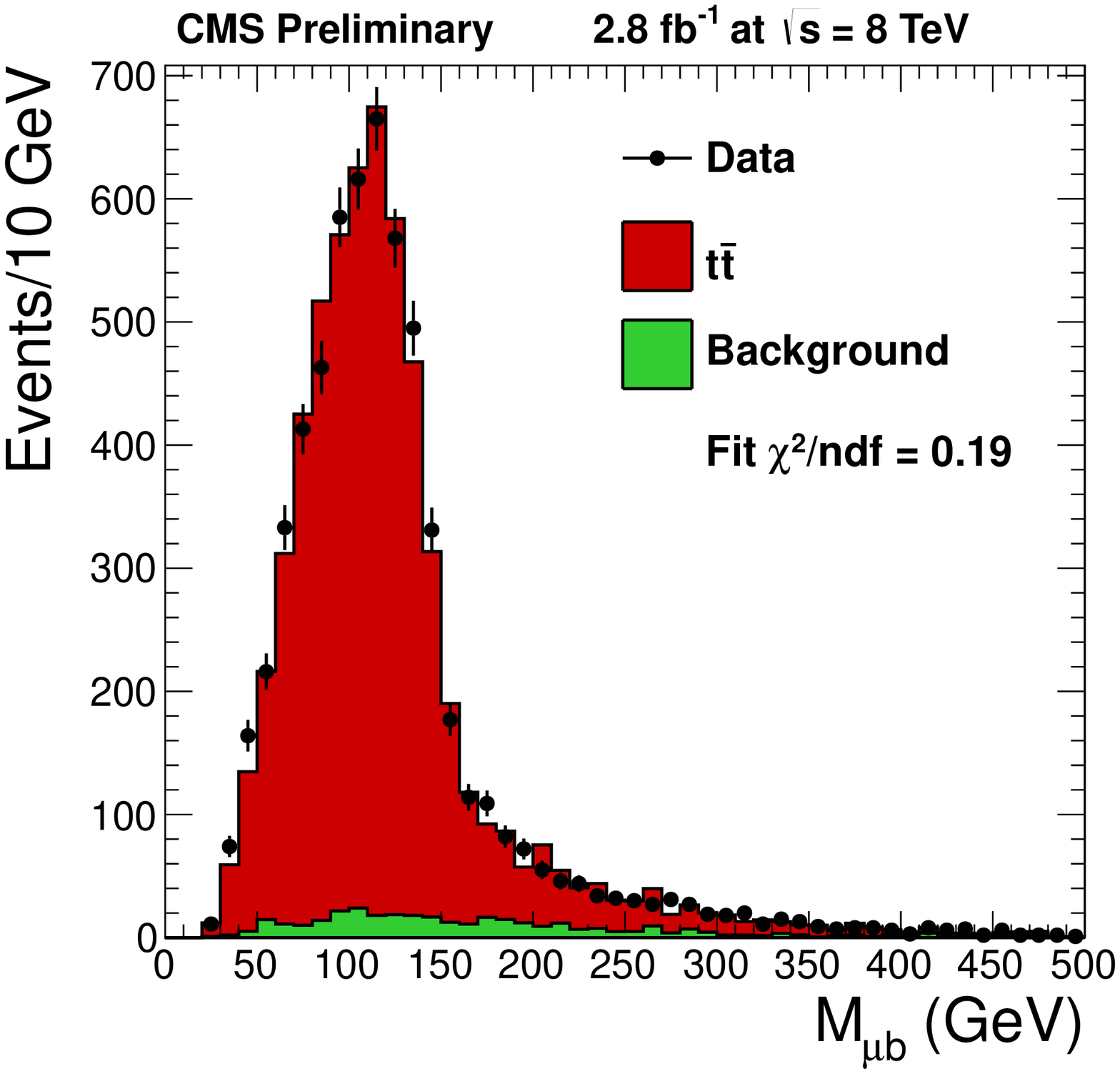}
\caption{\label{fig:cms_ljets_8tev} The invariant mass distribution $M_{\ell b}$ for a) electrons and b) muons, the expected \ttbar cross section is normalized to the measured cross section of 8.27 pb.}
\end{SCfigure}

Two CMS measurements are presented at the increased energy of $\sqrt{s} = 8$ TeV, one in the \ljets channel \cite{cms_ljets8tev} and one using events in the dilepton channel ($ee$, $\mu\mu$ and $e\mu$) \cite{cms_dilepton8tev}. In case of the \ljets channel events are required to have an isolated electron or muon with $p_T > 30\,(26)$ GeV and $|\eta| < 2.5\,(2.1)$ and at least four jets. The two leading jets are required to have $p_T > 45$ GeV, whereas the next-two leading jets are selected if $p_T > 35$ GeV. All jets are required to be within $|\eta|$ of $2.5$ and at least one jet needs to be $b$-tagged. The \ttbar cross section is determined from a template fit to the invariant mass distribution of the lepton and the $b$ quark $M_{\ell b}$ (see Figure \ref{fig:cms_ljets_8tev} a) and b)).
The total \ttbar cross section assuming $m_t = 172.5$ GeV is then measured to be \mbox{$\sigma_{\mm{tot}}(t\bar{t}) = 228.4 \pm 9(\mm{stat}) ^{+29.0}_{-26.0} (\mm{sys}) \pm 10 (\mm{lumi})$} pb. The dominating systematic uncertainties arise from the jet energy scale and the $b$-tagging efficiency measurement.
\begin{SCfigure}[][ht]
\centering
\includegraphics[width=0.35\textwidth]{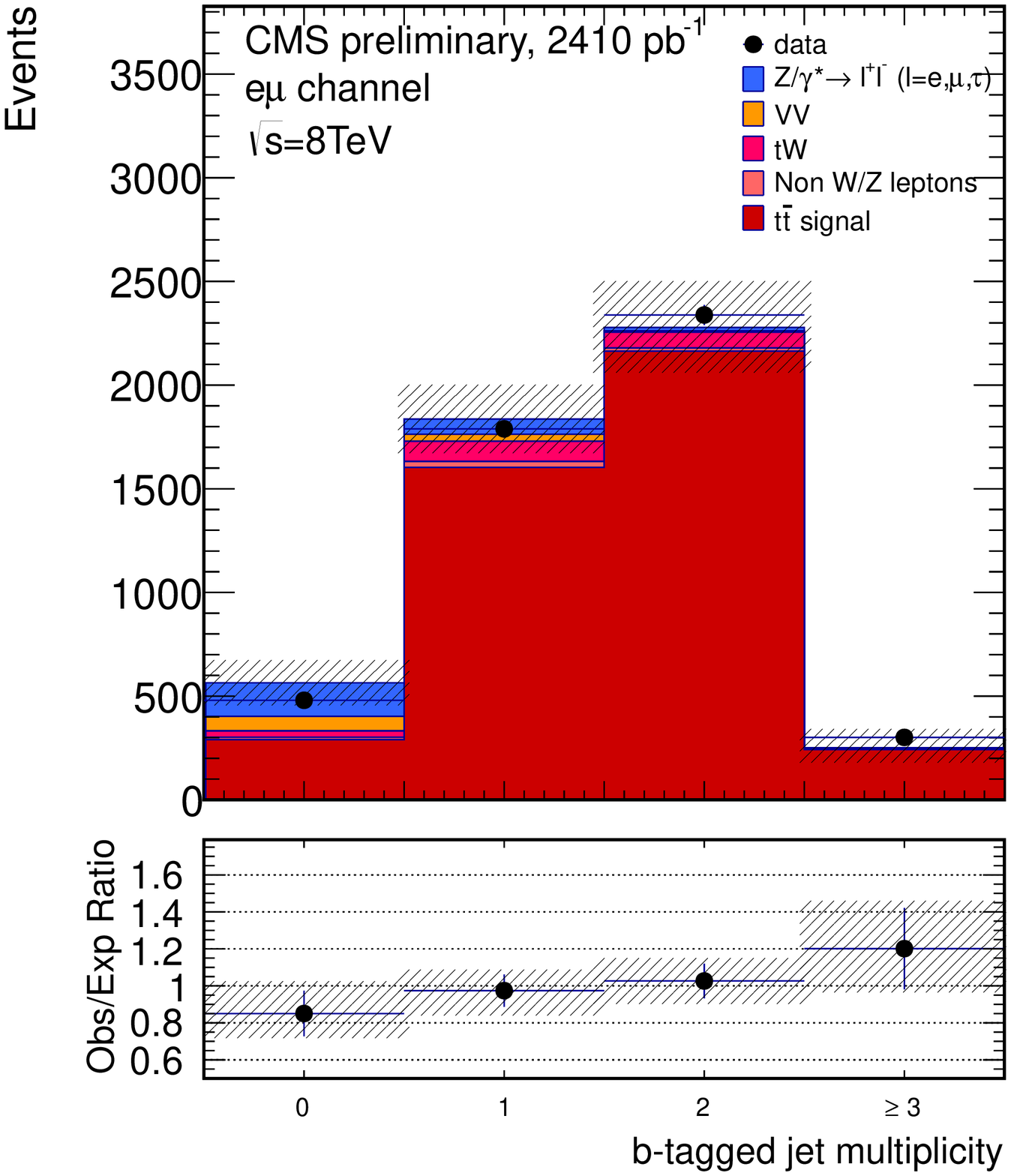}
\includegraphics[width=0.35\textwidth]{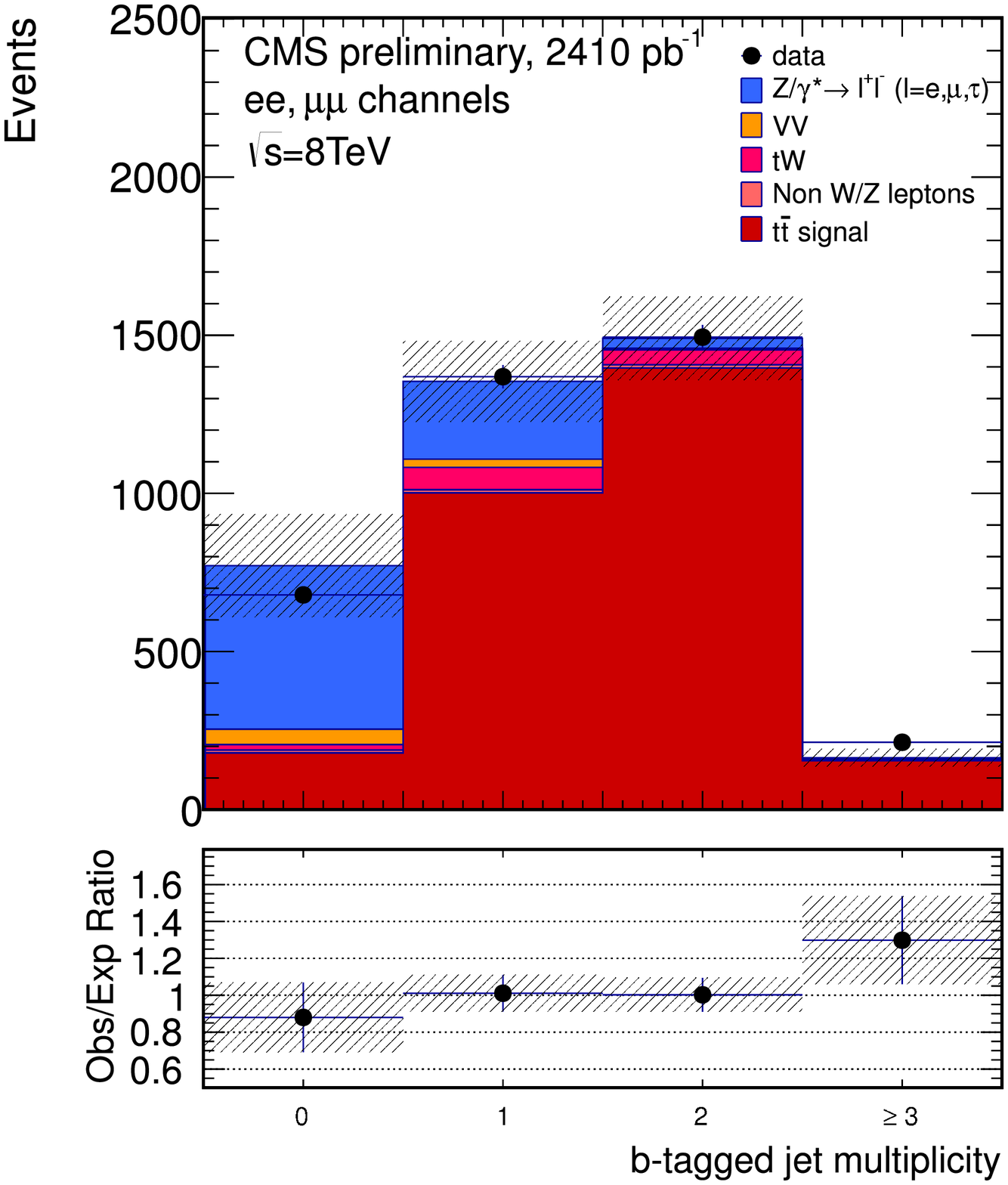}
\caption{\label{fig:cms_ll_8tev} Data are compared to MC for different number of $b$-tags for the a) $e\mu$ channel and for the b) summed $ee$ and $\mu\mu$ channel.}
\end{SCfigure}
The measurement is in agreement with the latest theory prediction of $\sigma_{\mm{tot}}(t\bar{t}) = 245.8 ^{+6.2}_{-8.4} (\mm{scales})~^{+6.2}_{-6.4} (\mm{pdf})$ pb \cite{nnloTheory}.\\
In addition a CMS measurement in the dilepton channel at 8 TeV is presented. Events are selected by requiring isolated electrons or muons with $p_T > 20$ GeV and $|\eta| < 2.5$ or $|\eta| < 2.4$, respectively. To suppress background from \zplus contributions an invariant mass window cut of $|m(\ell \ell) - m(Z)| > 15$ GeV for $ee$ and $\mu\mu$ events and in addition \met$> 40$ GeV is required. At least two jets with $p_T > 30$ GeV and $|\eta| < 2.5$ are required, of which at least one needs to be $b$-tagged. The low background contamination allows for a measurement of the \ttbar cross section using a counting method (see Figure \ref{fig:cms_ll_8tev}). The total \ttbar cross section assuming $m_t = 172.5$ GeV is measured to be $\sigma_{\mm{tot}}(t\bar{t}) = 227 \pm 3(\mm{stat}) \pm 11 (\mm{sys}) \pm 10 (\mm{lumi})$ pb, and the dominating systematic uncertainties arise from the jet energy scale, the trigger and lepton identification and isolation efficiencies. The measurement is in agreement with the latest theory prediction and other measurements in dilepton channel or by ATLAS (see Table \ref{tab:xsecLHCSummary}). Other top quark related physics results by ATLAS and CMS can be found here \cite{{LHCwebpages}}.

\begin {table}[htp]%
\centering %
\begin {tabular}{lclc}
\toprule %
Measurement & ${\mathscr{L}}$ [fb$^{-1}$] & $\sigma_{\mm{tot}}(t\bar{t})$ [pb] & total rel.\,unc. \\ \midrule
ATLAS 7 TeV (\ljets) & 0.7 & $179 \pm 4 (\mm{stat}) \pm 9 (\mm{sys}) \pm 7 (\mm{lumi})$ & 6.7\% \T \\
ATLAS 7 TeV ($\ell \ell$) & 0.7 & $176 \pm 5 (\mm{stat}) ~^{+14}_{-11} (\mm{sys}) \pm 8 (\mm{lumi})$ & 8.9\% \\
CMS 7 TeV (\ljets) & 2.3 & $158.1 \pm 2.1 (\mm{stat}) \pm 10.2 (\mm{sys}) \pm 3.5 (\mm{lumi})$ & 6.9\% \\
CMS 7 TeV ($\ell \ell$) & 2.3 & $162 \pm 2 (\mm{stat}) \pm 5 (\mm{sys}) \pm 4 (\mm{lumi})$ & 4.1\% \\ \\
Theory: & & & \\ 
NNLO pQCD & NA & $172.0^{+6.4}_{-7.5} (\mm{scales \oplus pdf})$ \cite{nnloTheory} & 4\% \T \\ \hline
ATLAS 8 TeV (\ljets) & 5.8 & $241 \pm 2 (\mm{stat}) \pm 31 (\mm{sys}) \pm 9 (\mm{lumi})$ & 13\% \T \\
ATLAS 8 TeV ($e\mu$) & 20.3 & $237.1 \pm 1.7 (\mm{stat}) \pm 7.4 (\mm{sys}) \pm 7.4 (\mm{lumi})$ & 4.7\% \\
CMS 8 TeV (\ljets) & 2.8 & $228.4 \pm 9 (\mm{stat}) ^{+29.0}_{-26.0} (\mm{sys}) \pm 10 (\mm{lumi})$ & 13\% \\
CMS 8 TeV ($\ell \ell$) & 2.4 & $227 \pm 3 (\mm{stat}) \pm 11 (\mm{sys}) \pm 10 (\mm{lumi})$ & 7\% \\ \\
Theory: & & & \\
NNLO pQCD & NA & $252.9 ^{+13.3}_{-14.5} (\mm{scales \oplus pdf \oplus \alpha_s})$ \cite{nnloTheory} & 5.5\%\\ \bottomrule
\end {tabular}
\caption {\label{tab:xsecLHCSummary} Summary of presented and discussed measurements of the total \ttbar production cross sections and their uncertainties at the LHC. The ATLAS measurement in the $e\mu$ channel has an additional uncertainty of $4.0$ pb originating from the uncertainty on the beam energy.}
\end {table}

\section{Conclusion}
Measurements of the \ttbar production cross section in electron and muon final states have been presented. The results from Tevatron and the LHC are in good agreement with theory predictions. More results from the Tevatron using the full data set are expected very soon. The results at the LHC are compatible between ATLAS and CMS and overall have small uncertainties despite the higher pile-up environment at $\sqrt{s}=8$ TeV compared to $\sqrt{s}=7$ TeV. The individual LHC measurements are challenging the precision of the theory calculations. The most precise measurement to date at $\sqrt{s}=8$ TeV uses events in the dilepton decay channel by ATLAS and more high precision results are to come.

\section*{Acknowledgments}
I would like to thank my colleagues from ATLAS, CDF, CMS and \dzero in preparing the presentation and these proceedings.
The author thanks the organizers of the TOP 2013 workshop for the invitation and for the hospitality of the conference venue.


\begin{footnotesize}

\end{footnotesize}


\end{document}